\PassOptionsToPackage{numbers,sort&compress}{natbib}

\documentclass{article}

\usepackage[preprint]{neurips_2026}

\usepackage[utf8]{inputenc}
\usepackage[T1]{fontenc}
\usepackage{hyperref}
\usepackage{url}
\usepackage{booktabs}
\usepackage{amsmath,amssymb}
\usepackage{nicefrac}
\usepackage{microtype}
\usepackage{xcolor}
\usepackage{graphicx}
\usepackage{float}
\usepackage{placeins}
\usepackage{newunicodechar}
\newunicodechar{≈}{\approx}
\usepackage{hyperref}
\hypersetup{colorlinks=true, linkcolor=blue, citecolor=blue, urlcolor=blue}

\title{Towards Efficient Auto-Interpretability of Boltz-1 Sparse Autoencoders}

\author{%
  Piotr Jedryszek\textsuperscript{1, 2}\thanks{Corresponding author: \texttt{piotrjedryszek@evolverebiosciences.com}. Affiliations: \textsuperscript{1}Dept.\ of Biology, Univ.\ of Oxford; \textsuperscript{2}Evolvere Biosciences, London; \textsuperscript{3}Kavli Institute, Univ.\ of Oxford; \textsuperscript{4}Dept.\ of Chemistry, Univ.\ of Oxford;}
  \textbf{Oliver M. Crook\textsuperscript{3, 4}}
}

\begin{document}
\maketitle

\begin{abstract}
Sparse autoencoders (SAEs) could turn AI models in Biology and other fields into engines of scientific discovery by explaining the superhuman capabilities of those models. We postulate that a latent is useful if we know three things: whether it is coherent, whether it can be described, and whether that description has predictive power. These questions are routinely conflated. We assemble them into a single pipeline and report some practical innovations. First, cross-seed dictionary stability prioritises which latents are worth spending resources to investigate. Second, an intruder-detection task asks whether a latent's activating examples share a recognizable pattern. Third, a separate pass proposes a candidate biological description which we convert into falsifiable predictions which can be tested in silico. Deployed on the Boltz-1 Pairformer trunk, stability prioritisation finds interpretable latents using about $4.4\times$ fewer latent evaluations each, and at $5.2\times$ lower measured LLM cost, while recovering over half of them. However, we notice that in our setup the LLM judge is heavily biased towards consistent residue identity and a simple filter that selects for features with high amino-acid identity precision outperforms the stability thresholding. This simpler method has the added benefit of requiring no extra SAE seeds but it remains uncertain if it reflects a genuinely useful heuristic or an limitation of the LLM judge. \footnote{\href{https://github.com/oxPJ/Efficient-Auto-Interpretability-of-AI-Models-in-Biology}{Code available here}}
\end{abstract}

\section{Introduction}

Models like AlphaFold~\cite{abramson_accurate_2024} and the open-source Boltz~\cite{wohlwend_boltz-1_2024} have allowed biologists to get accurate protein structure predictions in minutes rather than months. However, the inner workings of those models remain a mysterious and active field of research ~\cite{feldman_alphainterp_2026, gut_dissecting_2024, clore_explaining_2026, jedryszek_probing_2026}.

Sparse Autoencoders (SAEs) have emerged as powerful tools for dissecting neural-network activations into simpler, more interpretable units. While they tend to underperform supervised methods like linear probes~\citep{kantamneni_are_nodate}, they offer a path towards discovering new concepts due to their self-supervised methods of identifying simple components, known as features. Those features can then be explored by comparing example inputs on which the activations are or are not active. LLMs can be used to propose labels for those SAE features, a method especially useful in biology as the inputs and outputs of the model are not inherently interpretable. For this reason LLM auto-interp of neurons and SAE features has emerged as an exciting approach towards finding new scientific discoveries within Biological foundation models ~\citep{simon_interplm_2025, banerjee_automated_2025}.

However, deploying it on biological data exposes three separate problems. \textbf{(a) Cost.} Rigorous LLM evaluation across the ever greater number of latents SAE produce is not something all groups can afford. \textbf{(b) Interpretability is not the same thing as description quality.} The standard describe-then-simulate paradigm conflates an LLM's skill at writing explanations with the latent's intrinsic interpretability. We instead ask a question that needs no explanation and no predefined label. Given four windows where a latent fires and one where it does not, can a judge spot the odd one out \citep{paulo_evaluating_2025}? A single latent often has too few activating examples to train a classifier, but an LLM can still say whether a handful of examples share a recognisable pattern. \textbf{(c) There is no standard way to check whether a passing description is actually true.} Testing millions of latents means testing millions of hypotheses in a very high-dimensional space. A description can sound coherent while being verifiably false, for instance invoking a C-terminal motif for a feature also present at the N-terminus.

\textbf{The pipeline.} These problems become three areas we explore in sequence. (1) Cross-seed stability ranks latents before any LLM call is made. Given independently trained dictionaries latents shared across seeds tend to be easier to interpret \cite{paulo_sparse_2025} and steer\cite{jedryszek_stable_2026}. We find that it predicts interpretability in our set up, where it cuts latent evaluations per interpretable latent about $4.4\times$ while recovering just over half of the interpretable latents (Section~\ref{sec:stability}). However, we also find that a simpler signal: residue precision (the share of a latent's firing residues that are its most common amino acid), does at least as well on this panel. We suspect that this might be at least partly because in our setup the LLM judge largely focuses on residue identity (Section~\ref{sec:judge}). (2) Intruder detection tests whether a latent's activating examples are coherent, a task we adopt from \cite{paulo_evaluating_2025}. (3) A separate describe pass proposes a biological label, which we use to create to the falsifiable prediction it implies. We then formalise it in a simple Python script, which is used to validate the description using curated annotations (Section~\ref{sec:verify}). We observe that the selection in stage 1 (latent stability) keeps most latents whose activation tracks their residue across the corpus, and it discards most of those that fire on their residue only in specific contexts, including four of the five functional-site finds (of three zinc latents, a binding-site lysine and the kinase HRD histidine, it keeps only the lysine; Sections~\ref{sec:blindspot} and~\ref{sec:discussion}).

\section{Methods}
\label{sec:methods}

We analyze SAEs (TopK, $k = 256$, dictionary size $2048$, input dimension $384$) trained on Boltz-1 Pairformer trunk activations (layer 20, the middle of the 48-layer trunk, where we expected rich representations; recycle 1)\footnote{\href{https://huggingface.co/collections/evolve-away/boltz-saes}{publicly available Hugging Face collection of the SAEs}} and made available by a previous study \citep{jedryszek_probing_2026}. We target the trunk rather than the diffusion module because concurrent work indicates sequence-level biochemistry is linearly accessible there, and attenuated downstream~\citep{jedryszek_probing_2026}. Activations come from a 196,175-residue SwissProt subset, and a latent \emph{fires} on a residue when its activation reaches $0.6\times$ its maximum. Full architecture, density-prefiltering, and windowing details are in Appendix~\ref{sec:appendix_methods}.

\textbf{Calibrated intruder evaluation.} An LLM judge (Claude Sonnet 4.6, temperature 0) is shown four genuine activating windows plus one intruder window and must identify the intruder (construction in Appendix~\ref{sec:appendix_intruder}). Each window shows its highlighted residues, the protein name and up to five nearby SwissProt annotations, and the prompt asks the judge to start from residue identity (Appendix~\ref{sec:appendix_prompts}). We found the standard restricted-budget version of this task completely fails as a harness on ground-truth controls. Increasing the output-token allowance from 16 to 512 tokens restored accuracy on 20 amino-acid positive controls from 0.17 to 0.97, while prompt wording and metadata changed it far less, spanning 0.85 to 0.97 across four phrasings (Appendix~\ref{sec:appendix_robustness}).

\textbf{Description generation.} Concept labels used later (e.g.\ ``fires on zinc-coordination residues'', Section~\ref{sec:verify}) come from a separate \emph{describe} pass. An independent LLM call, the same judge model, is shown a feature's activating windows and asked for a one-sentence description. This pass is fully decoupled from intruder detection: the intruder judge never sees the description. The description exists only to attach a human-readable hypothesis to a latent that already passed, or failed, the blind intruder task. The describer also sees nearby SwissProt annotations (Appendix~\ref{sec:appendix_verify}).

\textbf{Ground-truth category matching.} For each description we identified, by hand and before computing any enrichment, the residue class it names and the SwissProt or DSSP category its context implies (e.g.\ ``zinc-coordination'' $\to$ \texttt{binding\_site:zn2} / \texttt{zinc\_finger}). We report every context fold-enrichment conditional on the named residue class, since an all-residue background over-credits a detector that already fires on the right residue, and each fold carries a by-protein bootstrap 95\% confidence interval because residues within a protein are not independent. Automating this process into an iterative loop is future work (Section~\ref{sec:discussion}).

\textbf{Cross-seed stability.} Following \cite{jedryszek_probing_2026} we deduplicate each seed's dictionary (merging directions above cosine 0.95), then match every alive seed-1 direction one-to-one to seeds 2 and 3 by decoder cosine (Hungarian assignment). A latent's stability is the mean of its matched absolute decoder cosine across the two held-out seeds, a continuous value in $[0,1]$; this is the axis in Fig.~\ref{fig:fig4_discovery} and the quantity we threshold at 0.5 (the 21-layer proxy skips the deduplication; Appendix~\ref{sec:appendix_method_details}). Separately, we call a latent \emph{shared} when, in both held-out seeds, its decoder and its encoder are matched to the same partner with $|\cos\theta|>0.7$; the per-concept recurrence rates use this binary flag.

\textbf{Supervised proxies and subspace stability.} Two LLM-free, label-based views of interpretability (12,204 alive latents across 21 layers scored against 15 curated concepts without amino-acid identity; 735 per-layer concept winners with it) and a subspace-level stability measure are detailed in Appendix~\ref{sec:appendix_method_details}; resampling details are in Appendix~\ref{sec:appendix_stats}. ``Low-frequency'' means the 20th to 60th activation-frequency percentiles (pooled across layers for the 21-layer proxy).

\textbf{Residue precision, recall and correlation.} With $X$ a latent's most common firing residue, its \emph{residue precision} is the share of its firing residues that are $X$, and its \emph{residue recall} the share of all $X$ on which it fires. Its \emph{residue correlation} $r$ is the correlation over the corpus between ``the residue is $X$'' and its activation, which also counts weak activations. All three need only the SAE and the sequences. A latent is \emph{pure} when its residue precision is $\geq 0.5$, given at least 20 firing residues.

\section{Results}

\subsection{Cross-seed stability predicts interpretability but not better than residue precision}
\label{sec:stability}

We asked whether cross-seed stability could help prioritise latents for LLM evaluation. We first tested its association with recovery of annotated concepts, then measured its usefulness in the LLM intruder test. Finally, we compared it with residue precision, a simpler signal requiring only one SAE.

\subsubsection*{Stability tracks recovery of known concepts}

We first assessed interpretability using annotated concepts (no LLM). For each concept in each of 21 layers, we selected the latent with the highest concept-recovery F1, yielding 735 concept winners. Among these winners, cross-seed stability correlated with F1 ($\rho = 0.51$, 95\% CI $0.39$ to $0.59$; Fig.~\ref{fig:fig3_prefilter}B).

However, this association partly reflected differences between concept types. Amino-acid identity winners had both high concept recovery and high reproducibility (mean F1 0.84; shared across seeds 56\% of the time). Secondary-structure winners were less reproducible (F1 0.67; shared 16\%), as were SwissProt functional-annotation winners (F1 0.21; shared 15\%). Within a concept type, the correlations were much weaker, ranging from $\rho = 0.09$ to $0.19$.

We next tested whether stability could enrich a broader candidate pool for latents recovering known concepts. This analysis scored all 12,204 alive latents using a concept benchmark that excluded amino-acid identity. Among the 4,881 latents in the discovery band (activation-frequency percentiles 20 to 60, pooled across layers), selecting those with stability $\geq 0.5$ nearly doubled the proportion meeting the concept-recovery criterion ($1.97\times$ enrichment; layer-clustered bootstrap 95\% CI $1.72$ to $2.24$; Fig.~\ref{fig:fig3_prefilter}A). Enrichment persisted when frequency percentiles were defined within each layer ($1.72\times$; additional frequency bands in Appendix~\ref{sec:appendix_band}).

\subsubsection*{Stability increases the yield of LLM evaluation}

We then tested stability as a filter for the LLM intruder test. We sampled 150 low-frequency latents from Boltz-1 layer 20, of which 141 passed density screening. This frequency band was selected after an earlier sampling design returned a null result (Appendix~\ref{sec:appendix_sweep}).

Of the 17 latents meeting the stability threshold of $\geq 0.5$, nine passed the LLM test (intruder accuracy $\geq 0.5$), giving a 53\% hit rate. These nine also represented 53\% of the 17 passing latents in the full panel (Fig.~\ref{fig:fig4_discovery}A). The hit rate was $4.4\times$ the panel-wide rate of 12.1\% (17/141), and $8.2\times$ the 6.5\% rate among unstable latents (Fisher OR $= 16.3$, $p = 8\times10^{-6}$).

Restricting evaluation to the stable subset reduced the number of evaluations per passing latent from 8.3 to 1.9, a $4.4$-fold reduction. Measured LLM cost per passing latent was $5.2$-fold lower, because the judge produced shorter responses for the selected latents. The reduction in evaluations per passing latent persisted across stability thresholds (Appendix~\ref{sec:appendix_thresholds}).

\subsubsection*{Residue precision provides a simpler baseline}

Amino-acid identity was the best-recovered and most reproducible concept type in the concept-winner analysis. The LLM judge also relied strongly on residue identity (Section~\ref{sec:judge}). These observations prompted us to test whether residue precision could provide a simpler way to prioritise latents.

We found that residue precision provided a better overall ranking than stability (Fig.~\ref{fig:free_triage}A): average precision was $0.77$ versus $0.40$ (difference $+0.36$, 95\% CI $+0.12$ to $+0.54$). At a budget of 17 evaluations, it recovered 12 passing latents versus 9 for stability (Appendix~\ref{sec:appendix_triage}).

A residue-precision threshold of $\geq 0.5$ selected 25 latents and recovered 14 of the 17 passing latents, requiring 1.8 evaluations per find. It retained four of the five functional-site finds, compared with one for stability. Adding stability to a logistic model based on residue precision did not significantly improve model fit (likelihood-ratio $p = 0.06$).

We note that these comparisons are exploratory and residue precision was proposed after observing its association with judge accuracy on this same panel.

\begin{figure}[t]
    \centering
    \includegraphics[width=0.9\textwidth]{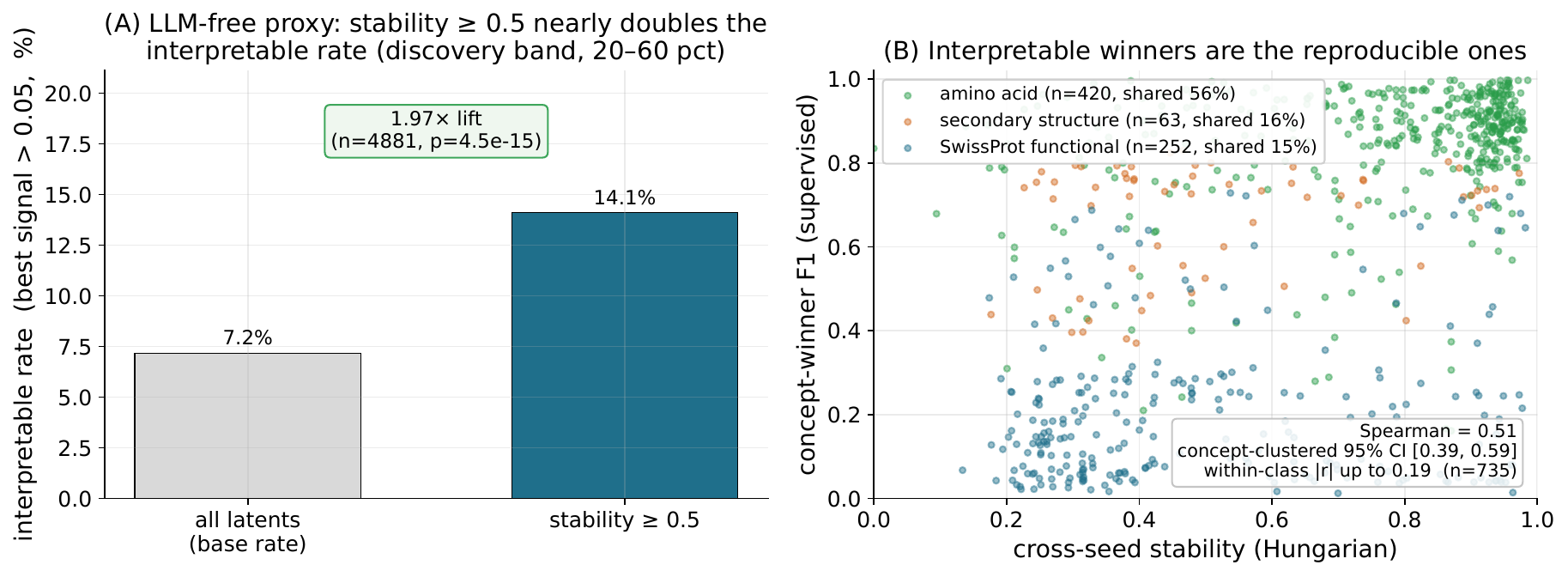}
    \caption{\textbf{Cross-seed stability as an LLM-free interpretability proxy.} \textbf{(A)} In the discovery band (activation-frequency percentiles 20 to 60, pooled across 21 layers), a stability $\geq 0.5$ filter nearly doubles the supervised interpretable rate (1.97$\times$, $n = 4{,}881$, layer-clustered 95\% CI $1.72$ to $2.24$); the LLM deployment also shows a lift (Fig.~\ref{fig:fig4_discovery}). \textbf{(B)} Concept-winner F1 vs.\ stability ($\rho = 0.51$, 95\% CI $0.39$ to $0.59$, $n = 735$). Amino-acid identities recur across seeds (56\%) far more than secondary structure (16\%) or SwissProt function (15\%).}
    \label{fig:fig3_prefilter}
\end{figure}

\begin{figure}[t]
    \centering
    \includegraphics[width=0.9\textwidth]{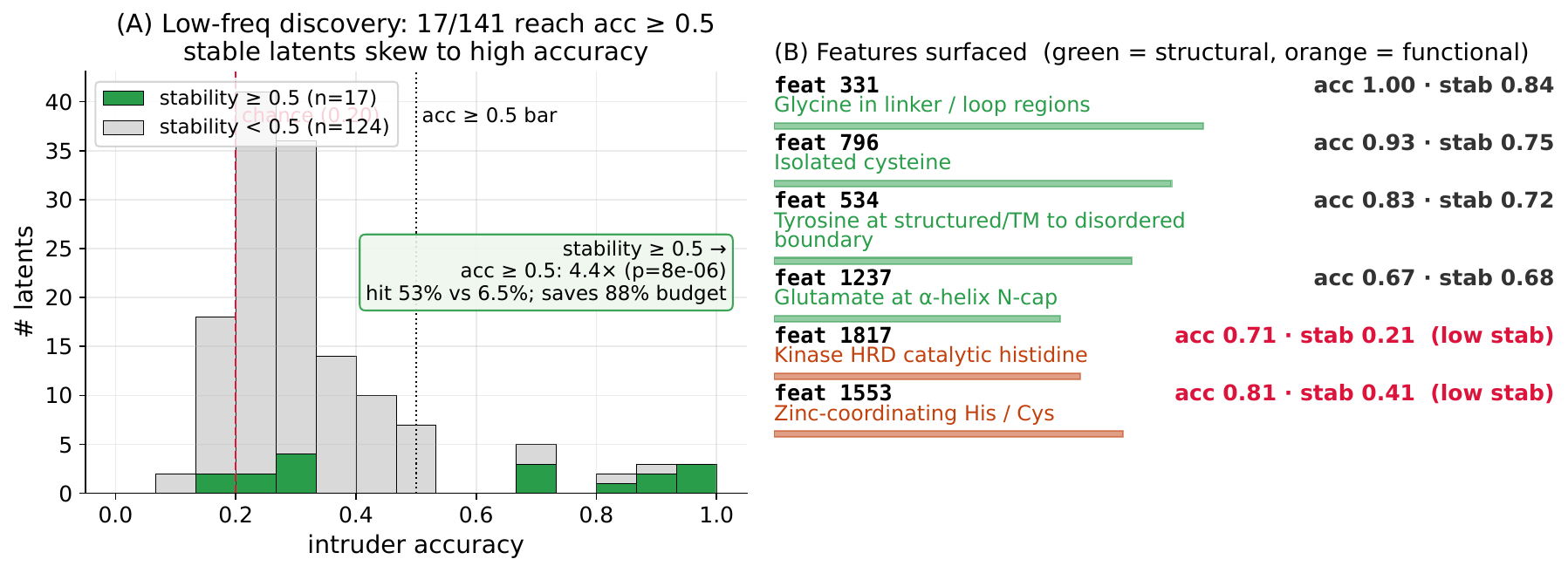}
    \caption{\textbf{Low-frequency discovery landscape.} \textbf{(A)} Distribution of intruder accuracy across the panel, stacked by stability band: judging only the 12\% of latents that clear stability $\geq 0.5$ costs about 4.4$\times$ fewer latent evaluations per interpretable latent and recovers 53\% (nine of 17) of them ($p = 8\times10^{-6}$). \textbf{(B)} Six of the discovered features by accuracy and stability. The functional motifs shown (catalytic histidine, metal coordination) fall at low stability, a pattern we examine in Section~\ref{sec:blindspot}.}
    \label{fig:fig4_discovery}
\end{figure}

\subsection{Intruder accuracy largely reflects residue identity}
\label{sec:judge}

The judge's accuracy increased with residue precision across the panel (Spearman $\rho = 0.52$, $p = 1\times10^{-10}$; Appendix Fig.~\ref{fig:fig5}A). Latents with residue precision $\geq 0.5$ had substantially higher odds of passing the interpretability test (odds ratio 140). 

To investigate why residue precision correlates with the LLM judge's accuracy, we asked how much intruder-detection performance could be reproduced using residue identity alone. We evaluated three rules based only on the highlighted residues: residue composition, composition across nine physico-chemical classes, and the absence of the most common residue. Across all 5,640 trials, their accuracies were $0.363$, $0.351$ and $0.326$, respectively, compared with $0.343$ for the LLM judge. The residue-composition rule achieved slightly higher accuracy than the LLM (difference $+0.021$, 95\% CI $0.006$ to $0.035$). Applying the same per-latent accuracy threshold of $\geq 0.5$, 16 of the 17 latents that passed the LLM test also passed this rule-based test. This overlap supports the interpretation that much of what the intruder evaluation finds is consistent residue identity. 

\begin{figure}[t]
    \centering
    \includegraphics[width=0.85\textwidth]{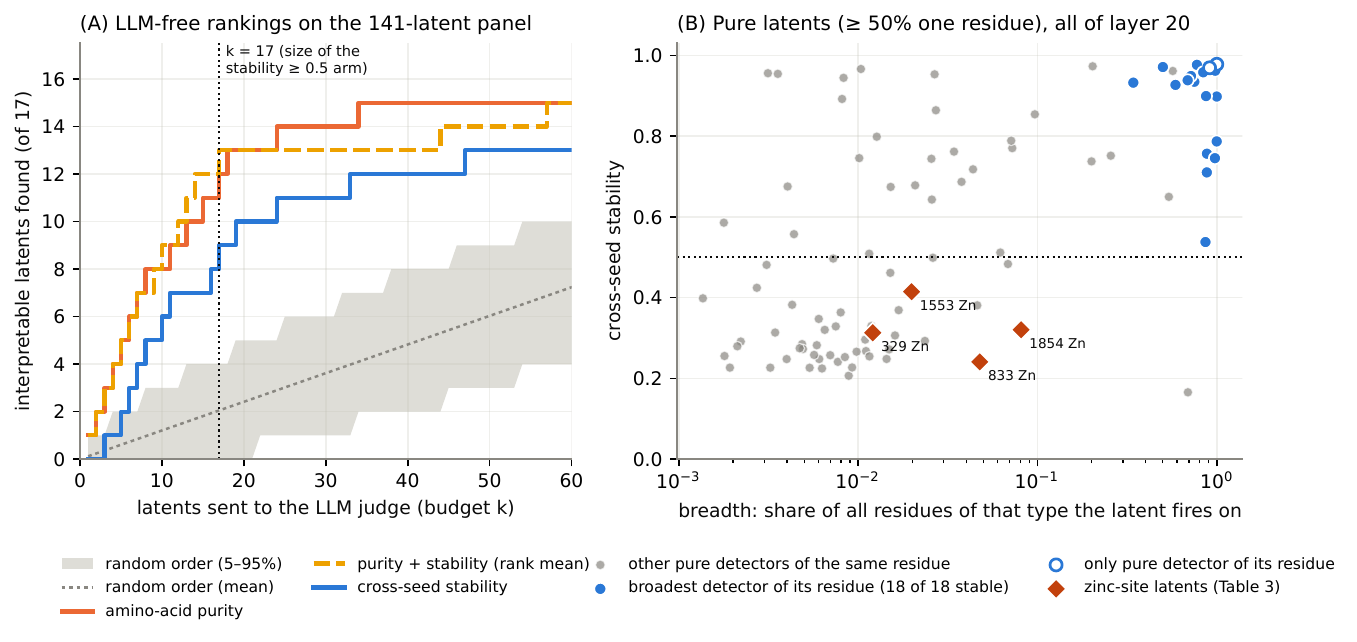}
    \caption{\textbf{Residue precision and stability as LLM-free triage signals.} \textbf{(A)} Latents passing the LLM test versus evaluation budget, ranked by residue precision, stability, or mean rank. Five latents with fewer than 20 firing residues are ranked last by residue precision. \textbf{(B)} Residue recall and stability for 98 pure layer-20 latents.}
    \label{fig:free_triage}
\end{figure}

\subsection{Auto-interp descriptions as falsifiable hypotheses}
\label{sec:verify}
\label{sec:two_kinds}

We tested the separately generated descriptions as biological hypotheses. For example, ``fires on zinc-coordination residues'' predicts enrichment at annotated zinc-binding sites. These checks assess the description's accuracy: a failed prediction does not establish that the latent is uninterpretable just that the description of evaluation was lacking.

We evaluated all 17 latents passing the intruder test (Table~\ref{tab:named_finds} in Appendix~\ref{sec:appendix_named_finds}). All residue claims validated (fold-enrichments $6.1$ to $84.9\times$), as expected given that the describer saw the highlighted residues. Context claims validated in 11 of the 16 testable cases. We measured context enrichment among residues of the named class, so residue identity alone could not explain it. Excluding proteins shown to the describer changed no verdict; broader exclusions left two latents untestable and a third below the validation threshold (Appendix~\ref{sec:appendix_verify}).

Residue correlation distinguished two kinds of detector (Appendix~\ref{sec:appendix_two_kinds}). Seven were \emph{graded} ($r = 0.53$ to $0.98$), activating strongly in some contexts and weakly on many other occurrences of their preferred residue. Ten were \emph{context-restricted} ($r = 0.02$ to $0.18$), covering at most 18\% of their preferred residue even at a lower activation threshold of 0.3 of their maximum. This second group included all five functional-site finds. Context claims validated for 9 of the 10 context-restricted detectors, compared with 2 of the 6 graded detectors with testable contexts (Fisher $p = 0.04$).

The most specific claims concerned zinc coordination and the HRD motif. Four zinc latents, including feature 329 below the intruder-test threshold, showed 6- to 13-fold enrichment among histidines and cysteines, with protein-bootstrap confidence intervals excluding 1 (Appendix Fig.~\ref{fig:fig5}B). The HRD-motif histidine latent detected nine of the 23 His-Arg-Asp motifs in the corpus, a 114-fold enrichment among histidines (protein-bootstrap 95\% CI 68 to 178). Within HRD-containing proteins, it fired only on the motif's histidine.

\subsection{Stability filtering misses validated motifs}
\label{sec:blindspot}

The four zinc-coordination latents and the HRD-motif histidine all fell below the stability cutoff of 0.5 (range $0.21$--$0.41$; Fig.~\ref{fig:fig4_discovery}B). Four passed the intruder test; the fifth, feature 329, scored $0.42$ but still validated against zinc annotations (Appendix Table~\ref{tab:named_finds}). More broadly, the filter retained 6 of the 7 graded finds but only 3 of the 10 context-restricted finds ($p = 0.05$).

We tested whether these motifs might recur across seeds through different bases of a shared subspace \citep{bhalla_sparse_2026}. Their mean subspace stability was $0.13$, compared with $0.36$ for amino-acid controls and no higher than for latents failing the intruder test (Appendix~\ref{sec:appendix_subspace_stability}). We therefore found no evidence of a shared subspace under the tested grouping, although the sample of five latents limits generalisation.

\section{Discussion}
\label{sec:discussion}

Our results support separating three decisions in auto-interpretability: which latents to evaluate, what the evaluation measures, and whether a generated description is biologically supported. Stability improved the efficiency of finding latents that passed the intruder test, but the judge largely responded to residue identity (Sections~\ref{sec:stability} and~\ref{sec:judge}). Residue identity is a legitimate interpretable property; success on this task nevertheless provides limited evidence that the judge recognises structural or functional context. Evaluations targeting contextual biology should therefore include intruders matched on residue identity.

Under this judge, residue precision achieved higher average precision than stability, required only one SAE, and retained more validated functional-site finds. This supports testing simple baselines before investing in additional dictionaries. Stability's measured LLM savings also exclude the training cost of the extra seeds, which would likely outweigh those savings at this panel size (Appendix~\ref{sec:appendix_cost}). Its practical value therefore depends on the evaluation budget and whether independently trained dictionaries already exist.

The validated motifs also caution against using stability as a hard exclusion rule (Section~\ref{sec:blindspot}). On this panel, the filter favoured graded residue detectors while excluding several context-restricted latents with supported biological associations. The subspace analysis did not recover cross-seed agreement under the tested grouping, but the sample of five motif latents limits generalisation. These findings also leave computational relevance unresolved: annotation enrichment establishes an association with biology, while determining whether Boltz-1 uses these features requires causal intervention.

Description validation provides a separate check of biological claims (Section~\ref{sec:verify}). Testing residue and context claims separately, with context enrichment measured against a residue-matched background, exposed unsupported context claims behind otherwise accurate residue descriptions. Failure to validate does not always refute a description, since some tests lack statistical power. The current mapping from descriptions to annotation categories also remains manual. Automating this step could make descriptions into systematically testable hypotheses, though claims beyond the available annotations would require additional evidence, potentially including wet lab experiments.

These conclusions are limited to one model, layer and judge, and 141 latents from a frequency band selected after an earlier null result. Residue precision and the detector groups were identified on this same panel, which was not assessed by human judges or evaluated without metadata. Held-out panels, additional models and judges, and human assessment are needed to establish how far these findings generalise.

\section*{Responsible Use Statement}
We analyse public Boltz-1 representations and SwissProt annotations and do not generate novel sequences, structures, or designs. The identified motifs are well characterised, so direct misuse risk is low. However, cheaper auto-interpretability could support harmful latent discovery in other models. We screened discovered descriptions for pathogenicity-related concepts and found none.

\section*{Acknowledgements and funding}

PJ is supported by the Biotechnology and Biological Sciences Research Council (UKRI-BBSRC) grant BB/T008784/1. OMC acknowledges funding from a New College Todd-Bird Junior Research Fellowship and MRC Fellowship MR/Y010078/1.

\bibliography{references}

\clearpage
\appendix

\section{Supplementary Methods}
\label{sec:appendix_methods}

\subsection{Architecture and Extraction}
TopK SAE, $k = 256$, dictionary size $2048$, input dimension $384$, $L_2$ regularizer weight $3\times10^{-3}$. Exemplars are compiled via a dual-pass script: Pass 1 computes global per-feature maxima and localized protein peaks; Pass 2 extracts a 33-residue window (radius 16) around each peak, binned into activation deciles: deciles 1 to 9 supply activating windows (up to 8 each) and decile 0 supplies negative proteins only.

\subsection{Method details}
\label{sec:appendix_method_details}
\textbf{Layers.} The LLM stages use layer 20 only, to keep their cost down. The supervised proxies span 21 Pairformer layers (every second layer from 0 to 40).

\textbf{Description generation.} The describer is shown a sample of a feature's activating windows grouped by activation strength, each with its protein name and up to six nearby SwissProt annotations (which can include SwissProt helix, strand and turn ranges), and asked for a one-sentence description. Because the describer sees SwissProt annotations for the windows it is shown, checking its description against SwissProt is not fully independent, so Appendix~\ref{sec:appendix_verify} also re-scores every claim with those proteins held out. The secondary-structure categories, which come from DSSP on Boltz-predicted structures, were never shown to it.

\textbf{Cross-seed stability.} The 21-layer proxy and the layer-wide analysis of Appendix~\ref{sec:appendix_stability_breadth} use the one-to-one matching of Section~\ref{sec:methods} without the deduplication step; on panel latents the stabilities with and without deduplication differ by at most $0.07$, which moves one find (feature 1499) from just above $0.5$ to just below. A max-cosine variant drops the one-to-one constraint, letting each latent take its best match in each held-out seed even when another latent took it too. Separately, we call a latent \emph{shared} when, in both held-out seeds, its decoder and its encoder are matched to the same partner with $|\cos\theta|>0.7$ on both; the per-concept recurrence rates use this binary flag.

\textbf{Residue precision, residue recall and residue correlation.} Residue precision and residue recall are the latent's precision and recall as a detector of its residue, the sense InterPLM's F1 uses \citep{simon_interplm_2025}; for the explanation ``fires on $X$'' the two swap, which is the sense of \citet{paulo_automatically_2024} and \citet{gao_scaling_2024}. The residue correlation is low when a latent fires strongly on little of its residue and its weaker activations are not specific to it.

\textbf{Supervised concept-recovery proxies.} We use two label-based views of interpretability, neither needing an LLM, both computed at recycle 1 across 21 Pairformer layers. The first view scores every alive latent (encoder-decoder cosine above 0.1), giving 12,204 latents. For each we take its best recovery across a 15-concept benchmark of three secondary-structure classes (AlphaFold-DSSP) and 12 SwissProt annotation categories (amino-acid identity is left out; full list in the next subsection). A concept's score is its F1 minus a per-concept shuffled-label null, and a latent counts as interpretable when its best score exceeds 0.05. The second view scores concepts: for each concept we take its single best latent per layer and read off that latent's cross-seed stability, now including the 20 amino-acid identities alongside the secondary-structure and SwissProt concepts, giving 735 concept winners (Fig.~\ref{fig:fig3_prefilter}B). Because these winners reuse the same concepts across layers, we summarise their F1 to stability relationship with a concept-clustered bootstrap 95\% confidence interval rather than a nominal $p$-value. Throughout, ``low-frequency'' means the 20th to 60th activation-frequency percentile band, the band the discovery sweep uses; for the 21-layer proxy these percentiles are pooled across layers.

\textbf{Subspace stability.} A genuinely multi-dimensional concept can be spanned by an arbitrarily rotated basis in each seed, which single-axis matching misses, so we also measure stability at the subspace level. In each seed we find the group of up to eight latents that co-activate with the target latent, meaning those whose per-residue firing correlates most strongly with it, and take the span of their decoder directions. Subspace stability is the overlap of these spans across seeds, computed as the mean $\cos^2$ of their principal angles. We validate this metric with positive controls (Appendix~\ref{sec:appendix_subspace_stability}).

\subsection{Benchmark concepts}
The supervised benchmark scores 15 concepts. The three secondary-structure classes are the AlphaFold-DSSP three-state labels (helix, sheet, coil). The 12 SwissProt categories are \texttt{helix}, \texttt{beta\_strand}, \texttt{turn}, \texttt{region:disordered}, \texttt{signal\_peptide}, \texttt{disulfide\_bond}, \texttt{binding\_site:substrate}, \texttt{modified\_residue:phosphoserine}, \texttt{glycosylation:n-linked\_glcnac\_asparagine}, and the three compositional-bias classes \texttt{basic\_and\_acidic\_residues}, \texttt{low\_complexity}, and \texttt{polar\_residues}. The 20 amino-acid identities enter only the concept-winner view (Fig.~\ref{fig:fig3_prefilter}B), not this 15-concept benchmark.

\subsection{Density Pre-filtering}
\begin{equation}
\text{Threshold}_{\text{scaled}} = 0.6 \times \text{Activation}_{\text{global\_max}}
\end{equation}
Latents with cumulative activation density $>0.85$ are flagged saturated and discarded, and a latent must fire in $\geq 10$ proteins to be \emph{sweep-eligible}. This exemplar-pipeline screen is what the 150-latent discovery sweep is drawn through, and it is separate from the \emph{alive} criterion used for the supervised proxy in Appendix~\ref{sec:appendix_method_details}, which is an encoder-decoder cosine above 0.1.

\subsection{Discovery sweep and intruder scoring}
\label{sec:appendix_sweep}
The discovery sweep samples 150 deduplicated layer-20, recycle-1 latents as three equal-count activation-frequency bins $\times$ 50, spanning the 20th to 60th activation-frequency percentiles; 141 clear the density/sparsity screen. Each latent is scored by intruder detection (4 genuine windows + 1 intruder, chance 0.20, construction in Appendix~\ref{sec:appendix_intruder}), with six trials per populated activation decile, giving 24 to 54 trials per latent (median 42). We report the mean accuracy per latent. Throughout, a latent counts as interpretable at intruder accuracy $\geq$ 0.5, a deliberately conservative bar: against chance 0.20 it corresponds to one-sided binomial $p \leq 10^{-3}$ even at the smallest trial count in the sweep. The released code also stores a binomial-test flag (\texttt{interpretable}, $p < 0.05$ against chance), a weaker criterion that marks 61 of the 141 latents; every result in this paper uses the $0.5$ bar.

\textbf{How the panel was chosen.} The low-frequency band was not our first design. An earlier comparison of the most and least stable latents, matched on activation frequency, found no difference, because frequency matching placed both arms in a high-frequency band where almost no latent is crisply interpretable. A broader sample did show a correlation, but it was driven partly by amino-acid latents being both stable and interpretable by construction. We then stratified by frequency and focused the sweep on the lower band, where the supervised proxy showed the stability enrichment to concentrate. Appendix~\ref{sec:appendix_band} shows that this band is not the only place the effect appears, but the panel's stability numbers should be read as coming from a band chosen with knowledge of where stability helps.

\subsection{Statistics}
\label{sec:appendix_stats}
All intervals are percentile bootstrap 95\% intervals except the two Mantel--Haenszel odds-ratio intervals (Appendices~\ref{sec:appendix_judge} and~\ref{sec:appendix_intruder}), which are analytic (Robins--Breslow--Greenland) intervals stratified by latent, and the Wilson interval for the judge's accuracy on the 150 identity-matched trials. Enrichment folds resample proteins (2,000 replicates), because residues within a protein are not independent. Panel statistics resample latents: 2,000 replicates for the residue-identity and mark-count rules, 10,000 for the paired comparison of rankings, and 20,000 for the cost ratio. The 21-layer supervised lift resamples whole layers (2,000 replicates), and the concept-winner correlation resamples concepts together with all their layers (5,000 replicates).

\subsection{Where the supervised lift sits}
\label{sec:appendix_band}
The 1.97$\times$ lift in Section~\ref{sec:stability} uses activation-frequency percentiles pooled across all 21 layers. Its layer-clustered bootstrap 95\% interval is $1.72$ to $2.24$, so the result does not rest on treating latents from different layers as independent. Pooled percentiles mix layers unevenly, because layers differ in how often their latents fire. Percentiles computed within each layer are a closer analogue of the LLM panel's band, though not the same set: on this proxy's layer-20 population the panel's latents sit at the 0th to 52nd percentiles. With within-layer percentiles the lift in the 20th to 60th percentile band is $1.72\times$ ($1.44$ to $2.02$); it persists into the 60th to 80th percentiles ($1.47\times$ and $1.44\times$ for the two deciles, both intervals above 1) and is absent in the top fifth ($0.97\times$ and $1.02\times$). Pooled banding gives $1.88\times$ ($1.35$ to $2.52$) below the 20th percentile and no lift above the 60th ($1.08\times$ and $1.03\times$ for the two upper quintiles, both intervals including 1). Across all 12,204 latents the lift is $1.47\times$ ($1.29$ to $1.67$).

\subsection{Cost accounting}
\label{sec:appendix_cost}
Every intruder trial is a separate request, submitted through the Anthropic Message Batches API (which prices at half the synchronous rate), so the panel's LLM bill is set by the trial count and not by the latent count. We quote the saving first in \emph{latent evaluations}, the unit the filter operates on: 1.9 evaluations per interpretable latent found under the filter against 8.3 unfiltered, a $4.4\times$ saving. The request ratio is effectively the same ($76.0$ against $331.8$ intruder requests per find, $4.4\times$), because the number of trials a latent receives does not depend on its stability (mean 40.2 in the stable arm against 40.0 elsewhere, Mann-Whitney $p = 0.84$); one evaluation is $40.0$ requests on average.

The measured dollar saving is larger, at $5.2\times$ (\$0.33 per find against \$1.69). Cost per \emph{trial} is not independent of stability, even though the trial count is: the stable arm averages $351$ output tokens per trial against $473$ elsewhere ($p = 6\times10^{-4}$), because a simple residue detector draws a short confident answer from the judge while an ambiguous latent deliberates to the token cap. The filter therefore retains the cheap latents as well as the productive ones, and the saving in money exceeds the saving in calls. We report both rather than collapsing them. Note that the ratio of \emph{total} spend (\$2.94 against \$28.76) is not a saving, since the filtered arm also returns fewer finds.

All token counts and dollar figures in Table~\ref{tab:cost} are measured, read back per request from the Batches API after the run, not estimated from prompt lengths. Our own earlier estimate, which priced output from the mean $839$-character judge transcript of the one run that persisted transcripts, understated output by $2.2\times$: that run covered high-interpretability latents, which answer tersely, while the panel is dominated by ambiguous latents that run to the cap. Since output is priced five times input, an estimate calibrated on the easy tail is not safe for a full panel.

\begin{table}[H]
\centering
\caption{\textbf{Measured cost of the 141-latent discovery sweep}, and of the stable-only arm the filter would have judged instead. Every figure is measured, not estimated: requests are exact, and token counts are read back per request from the Message Batches API after the run. Dollar figures use Claude Sonnet 4.6 batch pricing (half the synchronous rate). Token counts and dollar figures cover the intruder requests only. The describe pass is one request per latent, ran before the intruder trials were re-run with the current construction (descriptions do not depend on it), and is not included; nor is the describe module's activation-prediction step, one further request per latent that no result in this paper uses. Cost per find is the efficiency number; the ratio of total spend is not, because the filtered arm also returns fewer finds.}
\label{tab:cost}
\vskip 0.05in
\small
\begin{tabular}{lrr}
\toprule
 & \textbf{Stable-only ($n = 17$)} & \textbf{Full panel ($n = 141$)} \\
\midrule
Intruder requests            & 684              & 5{,}640 \\
Describe requests            & 17               & 141 \\
Input tokens                 & 0.76\,M          & 6.24\,M \\
Output tokens                & 0.24\,M          & 2.59\,M \\
Batch cost                   & \$2.94           & \$28.76 \\
\midrule
Interpretable latents found  & 9                & 17 \\
Intruder requests per find   & 76.0             & 331.8 \\
Cost per find                & \$0.33           & \$1.69 \\
\bottomrule
\end{tabular}
\end{table}

Two caveats belong with this number. First, the ratio itself is uncertain at this sample size: a bootstrap over latents gives a 95\% CI of $2.2$ to $8.5$ on the request-cost saving. Second, stability is free of \emph{LLM} cost but not of all cost, because it presupposes independently trained dictionaries. The marginal cost is smaller than it first appears, since additional seeds train on the same cached activations and require no further forward passes of the underlying model; but for a panel this size it is unlikely to amortise. Judging all 141 latents costs \$28.76, which two extra SAE training runs would likely exceed. The trade becomes favourable at the scale where auto-interpretability is actually expensive: judging a full $2{,}048$-latent dictionary at the same measured rate would cost roughly \$420, or \$43 under the filter. Groups without existing dictionaries should weigh SAE training against the evaluation budget they expect to save, rather than treating the prefilter as free. The residue precision $\geq 0.5$ arm avoids this cost entirely: it would judge 25 latents and find 14, or 1.8 latent evaluations per find, using only the one dictionary (we did not measure its dollar cost separately). We did not record the compute used to train the SAEs, so the break-even point above is an estimate rather than a measurement.

\subsection{Sensitivity to the 0.5 thresholds}
\label{sec:appendix_thresholds}
\begin{figure}[H]
    \centering
    \includegraphics[width=0.95\textwidth]{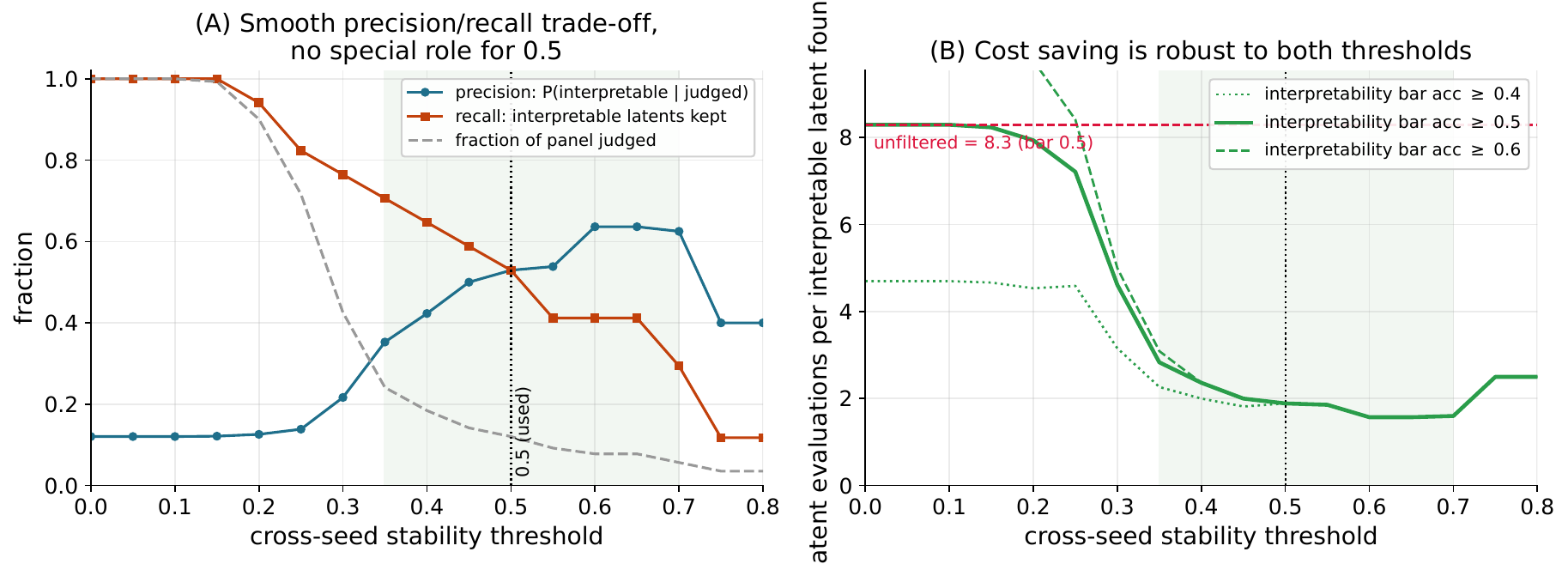}
    \caption{\textbf{Sensitivity to the stability cut and the interpretability bar.} \textbf{(A)} Precision, recall and the fraction of the panel judged, as a function of the stability threshold, at the 0.5 interpretability bar. The shaded band marks the 0.35 to 0.70 range quoted in the text and the dotted line marks the 0.5 cut used. \textbf{(B)} Latent evaluations per interpretable latent found, at interpretability bars of 0.4, 0.5 and 0.6. The saving over the unfiltered panel holds under all three.}
    \label{fig:figS4_thresholds}
\end{figure}

Here a filter's precision is its hit rate among the latents it sends to the judge, and its recall the share of the 17 finds it recovers. Both the stability cut and the interpretability bar are round numbers fixed in advance rather than tuned. Fig.~\ref{fig:figS4_thresholds} sweeps the stability cut across its range at three interpretability bars. Precision and recall trade off smoothly and monotonically, with no discontinuity at 0.5, and 0.5 is not an optimum. A cut of 0.60 gives higher precision (0.64) and lower cost (1.6 evaluations per interpretable latent) at lower recall (0.41). The conclusion is stable across a wide band: for any cut between 0.35 and 0.70 on a 0.05 grid, precision runs 0.35 to 0.64, recall 0.29 to 0.71, and cost 1.6 to 2.8 latent evaluations per interpretable latent against 8.3 unfiltered (2.9 at the worst cut between grid points). Two boundaries are worth stating. Below about 0.25 the filter does almost nothing, retaining 72 to 100\% of the panel at a lift of at most 1.15. Above 0.75 fewer than six latents survive and the estimates become unreliable.

\subsection{How an intruder trial is constructed}
\label{sec:appendix_intruder}
Each trial contains five 33-residue windows. The four genuine windows are drawn without replacement from a single activation decile of the target latent, each centred on that latent's activation peak in one protein. Residues firing above $0.6\times$ the window peak are marked inline and also listed explicitly to the judge. The intruder is a window centred on a \emph{different} latent's activation peak, drawn from a protein that contributes none of the target's own selected windows. Its highlights are then replaced by a genuine activation mask of the \emph{target} latent, taken from one of the target's own windows that is not shown in this trial. The five windows are shuffled and the judge, which sees no activation values and no feature description, must name the odd one out.

\textbf{The highlight positions carry almost no cue.} Transplanting a real mask of the target draws the intruder's mark count, run-length distribution and offset from the window centre from the target's own windows, which removes the centre cue. An adversary that sees the mark positions and nothing else scores $0.204$ across all $5{,}640$ trials of the sweep, against a chance of $0.200$. This matters because the obvious simpler choice does leave a usable cue. If only the number of highlighted residues is matched and the positions are drawn at random, genuine windows remain peak-centred while the intruder's marks do not, and the same adversary reaches $0.804$ by picking the window whose centre residue is unmarked. One small cue remains: the donor mask may come from a different activation decile than the four genuine windows, so the number of highlighted residues can differ, and a rule that picks the window whose mark count is furthest from the median of the five scores $0.242$ (latent-clustered 95\% CI $0.229$ to $0.254$). The judge picks up this cue. It is more accurate on the 898 trials where the intruder's mark count is the unique outlier than on the rest ($0.48$ against $0.32$), and the difference holds within latents (Mantel--Haenszel OR $2.1$, 95\% CI $1.8$ to $2.5$), while the identity-only rules of Section~\ref{sec:judge} are less accurate on those trials. The cue adds about $0.03$ to the judge's pooled accuracy of $0.343$, and on the trials without it the strongest identity rule leads the judge by more ($0.371$ against $0.317$). The residues underneath the marks are untouched, so their identities still differ between genuine and intruder windows; Section~\ref{sec:judge} shows how much of the judge's accuracy that identity signal accounts for.

\textbf{The intruder is not guaranteed silent.} The construction requires only that the intruder's protein contributes none of the target's own windows, and the target does fire somewhere inside the intruder window in $5.8\%$ of trials (329 of 5,640), concentrated in the densest latents (Spearman $\rho = 0.79$ against activation density). The effect on accuracy is negligible. Contamination is uncorrelated with per-latent judge accuracy ($\rho = -0.03$, $p = 0.70$), discarding every contaminated trial moves pooled trial accuracy from $0.343$ to $0.340$, and contaminated and clean trials do not differ significantly ($0.383$ against $0.340$, Fisher OR $1.2$, $p = 0.12$). One latent changes side of the interpretability bar: feature 915, contaminated in 22 of its 24 trials, reaches $0.5$ on its two clean trials. An earlier version of this analysis joined the judge's answers to trials rebuilt with a superseded intruder construction and reported a larger, significant gap; the numbers above use the trials the judge actually saw.

\textbf{Answer position.} The judge's accuracy rises with the intruder's position in the list, from $0.28$ when it is shown first to $0.40$ when it is shown last ($\chi^2 = 41.8$, $p = 2\times10^{-8}$). Positions are assigned by random permutation, so this bias adds noise but cannot favour particular latents.

\section{Supplementary Results}
\label{sec:appendix_results}

\subsection{What the judge tracks}
\label{sec:appendix_judge}
Residue precision is computed over the whole corpus, but the judge sees the residues of its five windows directly, so the two are less independent than they first look. On this panel, intruder detection passes most residue detectors (14 of the 25 pure panel latents) and gives no evidence that it reads context, and our production prompt, which tells the judge to start from the literal amino-acid identity, probably strengthens this. In the 150 trials whose intruder's most common highlighted residue matches the genuine windows', identity-only rules score $0.01$ to $0.02$; the judge's $0.153$, against $0.346$ on the other trials of the same 93 latents, gives a Mantel--Haenszel OR of $0.32$ (95\% CI $0.20$ to $0.50$). The judge does not simply follow the identity rule there, but it does no better than guessing.

\textbf{Residue precision, not residue recall.} The intruder differs from the genuine windows in which residues are highlighted, so nothing in the task checks whether a latent is silent on other residues of its type, and passing does not require it. Residue recall still helps a latent pass, but not because the judge checks silence. Among the 25 pure panel latents, accuracy rises with residue recall (Spearman $\rho = 0.77$), and the residue-composition rule, which by construction knows nothing about where a latent is silent, rises with it almost as much ($\rho = 0.72$). Part of the reason is that the task draws windows from every activation decile, and higher-recall detectors look more precise in them: the share of highlighted residues in the genuine windows that match their dominant residue tracks residue recall among the pure latents ($\rho = 0.69$; $0.26$ across all 136). This shown precision accounts for most of the rule's dependence on residue recall (partial $\rho = 0.31$ given it) but less of the judge's (partial $\rho = 0.50$, $p = 0.01$), which we cannot fully explain.

\subsection{Residue precision against stability as triage}
\label{sec:appendix_triage}
At the budget of 17 judgements, ranking by residue precision rather than by stability raises the hit rate by $+0.18$ (paired bootstrap 95\% CI $-0.06$ to $+0.41$), so of the two ranking comparisons in Section~\ref{sec:stability}, only the difference in average precision is significant. The mean of the two ranks has average precision 0.76 and finds 13 rather than 12 at that budget. The likelihood-ratio test of Section~\ref{sec:stability} compares logistic models of passing (intruder accuracy $\geq 0.5$), with and without stability, on the 136 latents that have a residue precision. Judging only the 10 latents that are both stable and pure gives 9 finds, a higher hit rate than either signal alone but 5 fewer finds than residue precision.

\subsection{Two kinds of residue detector: robustness and simulation scores}
\label{sec:appendix_two_kinds}
The split in Table~\ref{tab:two_kinds} uses $X$, the single most common firing residue, and with it no find falls between the groups ($r = 0.18$ to $0.53$). Scored on the residue classes the descriptions name, 1242 (Leu/Ile, $r = 0.42$) and 1805 (Ser/Thr, $0.23$) do. The same split appears in residue recall at 0.3 of the maximum (at most 0.18 against at least 0.28) and in InterPLM's best F1, an F1 above $0.5$ at the best of five activation thresholds (at most 0.21 against at least 0.34); InterPLM's criterion for amino-acid features, under which SAEs on even a weight-shuffled ESM-2 yield hundreds of them \citep{simon_interplm_2025}, would count six of our seven graded finds and none of the context-restricted ones. The difference between the kinds lies in how specific their weaker activations are. At 0.15 of its maximum, the glutamine latent 484 fires on half of all glutamines, as the leucine latent 1242 does on half of all leucines, but only 8\% of 484's firing at that level is on glutamine, against 67\% of 1242's on leucine. The zinc latent 329, just below the interpretability bar, is context-restricted too (residue correlation $0.07$).

\textbf{Simulation scores.} A perfect residue-only simulator, predicting 1 on every $X$ and 0 elsewhere, scored on random 64-residue windows alone gives a median correlation of $0.08$ on the context-restricted finds and $0.68$ on the graded ones. Under the top-and-random scoring of \citet{bills_language_2023}, five top-activating and five random windows, it scores $0.21$ to $0.75$ on the context-restricted finds (median $0.47$) against $0.52$ to $0.96$ on the graded ones, against an average of $0.15$ for GPT-4's explanations of GPT-2 neurons. A simulation score sees the difference only if it samples enough sequence in which the residue occurs and the latent is silent.

\begin{table}[H]
\centering
\caption{\textbf{The 17 finds: two kinds of residue detector, the annotation check's verdict on each context claim, and whether the stability filter kept each.} $X$: the latent's most common firing residue. Residue precision and recall at the firing threshold (0.6 of the latent's maximum), and recall also at 0.3 of it; precision is --- for 1817 and 1448, which fire on fewer than 20 residues, all on $X$. $r$: residue correlation. \emph{Context}: annotation-check verdict on the context claim (--- for 1652, whose description names no context; Section~\ref{sec:verify}; Appendix Table~\ref{tab:named_finds}). \emph{Kept}: stability $\geq 0.5$ (Section~\ref{sec:stability}).}
\label{tab:two_kinds}
\footnotesize
\begin{tabular}{rclccccll}
\toprule
 & & & \multicolumn{3}{c}{Residue precision and recall} & & \multicolumn{2}{c}{Check and filter} \\
\cmidrule(lr){4-6} \cmidrule(lr){8-9}
Feat & $X$ & Kind & Prec.\ (0.6) & Recall (0.6) & Recall (0.3) & $r$ & Context & Kept \\
\midrule
1237 & E & context-restricted & 1.00 & 0.004 & 0.04 & 0.02 & fails & yes \\
1817 & H & context-restricted & --- & 0.003 & 0.04 & 0.02 & holds & no \\
1553 & H & context-restricted & 0.58 & 0.020 & 0.05 & 0.04 & holds & no \\
1446 & K & context-restricted & 0.36 & 0.005 & 0.10 & 0.05 & holds & no \\
1499 & K & context-restricted & 0.70 & 0.007 & 0.06 & 0.10 & holds & yes \\
833 & C & context-restricted & 0.99 & 0.048 & 0.13 & 0.10 & holds & no \\
1805 & T & context-restricted & 0.57 & 0.021 & 0.06 & 0.13 & holds & yes \\
1448 & N & context-restricted & --- & 0.001 & 0.03 & 0.14 & holds & no \\
484 & Q & context-restricted & 0.65 & 0.008 & 0.18 & 0.15 & holds & no \\
1854 & C & context-restricted & 0.84 & 0.081 & 0.15 & 0.18 & holds & no \\
\midrule
1095 & F & graded & 0.51 & 0.068 & 0.46 & 0.53 & fails & no \\
1242 & L & graded & 1.00 & 0.027 & 0.28 & 0.55 & fails & yes \\
534 & Y & graded & 1.00 & 0.044 & 0.75 & 0.60 & fails & yes \\
331 & G & graded & 0.98 & 0.097 & 0.52 & 0.69 & holds & yes \\
796 & C & graded & 1.00 & 0.256 & 0.68 & 0.80 & holds & yes \\
1652 & W & graded & 1.00 & 0.977 & 1.00 & 0.95 & --- & yes \\
354 & P & graded & 1.00 & 0.865 & 1.00 & 0.98 & fails & yes \\
\bottomrule
\end{tabular}
\end{table}

\subsection{Annotation check: residue-conditional backgrounds and held-out proteins}
\label{sec:appendix_verify}
Conditioning every context fold on the named residue class changes one verdict: feature 354's context claim can barely beat the proline background (fold $0.94$) and now fails. Feature 331's glycine-in-coil fold falls from $1.85\times$ to $1.13\times$ but still holds, since most of the original fold was glycine's own preference for coil. The context is the half of a label that intruder detection, as deployed here, does not check (Section~\ref{sec:judge}).

The describe pass saw nearby SwissProt annotations for the windows it was shown, so we re-scored every claim, on the backgrounds of Table~\ref{tab:named_finds}, after removing those proteins from the corpus. No verdict changes. Removing every protein that contributes any of a latent's windows, shown or not, is stricter and costs power. Feature 1448 is left with no firing residues, so neither of its claims can be tested, and 1817 with two, neither on an HRD histidine (formally a fold of 0, but far too few residues to test its context claim). Feature 484's glutamine-in-disorder claim falls to $2.23\times$ (95\% CI $0.82$ to $3.60$, on 29 firing glutamines) and no longer clears the bar. Every other verdict is unchanged. One case shows why the check is not fully independent of what the describer saw: next to the HRD histidine, the describer was shown the kinase's annotated catalytic residue (``Active site: proton acceptor'', on the neighbouring aspartate). The HRD claim is scored against the His-Arg-Asp sequence motif rather than that annotation, but the hint was there.

\subsection{Stability and residue recall across layer 20}
\label{sec:appendix_stability_breadth}
Residue precision and recall show where stability is predictable. Across all 743 alive layer-20 latents that fire on at least 20 residues, stability and residue precision are not rank-correlated ($\rho = 0.02$), yet pure latents are three times as likely to be stable as the rest (50\% against 17\%; Fisher OR $5.0$, $p = 4\times10^{-12}$; unchanged after adjusting for training activation frequency). Within the 98 pure latents, what predicts stability is residue recall, the share of its residue type a latent fires on ($\rho = 0.58$, $p = 6\times10^{-10}$; partial $\rho = 0.58$ given training activation frequency). For every one of the 18 amino acids with more than one pure detector, the highest-recall detector is stable, against 10.0 expected if stability were shuffled among the detectors of each residue ($p = 2.5\times10^{-6}$); the most frequently firing detector is stable for only 12 (Fig.~\ref{fig:free_triage}B). The other pure detectors of the same residues are stable in only 29 of 78 cases, and the four zinc latents are among the unstable ones. Seeds always agree on the highest-recall detector of each residue, which covers more than a third of its residue at our firing threshold (34 to 100\%), the kind of amino-acid feature that SAEs find even in a weight-shuffled ESM-2 (Appendix~\ref{sec:appendix_two_kinds}).

This is not an artefact of one-to-one matching, which could penalise a concept that one seed splits into several latents and another does not. Dropping the constraint leaves the lowest-recall third of pure latents at a median stability of $0.33$ ($0.31$ with matching), and only 4 of these 33 reach a mean best-match cosine of $0.7$ in the other seeds. The zinc latents lie in the middle and highest-recall thirds rather than the lowest, and stay at $0.24$ to $0.42$ without the constraint. The subspace analysis gives the same answer for the five motifs (Appendix~\ref{sec:appendix_subspace_stability}, Fig.~\ref{fig:figS2_manifold_subspaces}): they score $0.13$, against $0.36$ for amino-acid controls run through the same firing-matched pipeline.

The two kinds also summarise which of the panel's finds the filter kept: 6 of the 7 graded finds and 3 of the 10 context-restricted ones (Section~\ref{sec:blindspot}). As continuous measures, residue recall at the firing threshold and the residue correlation track panel stability about equally across the 17 finds ($\rho = 0.52$ and $0.54$, both $p = 0.03$). Layer-wide, among the 98 pure latents, stability tracks the residue correlation ($\rho = 0.68$) at least as closely as residue recall at the firing threshold ($\rho = 0.58$; bootstrap 95\% CI of the difference $-0.004$ to $0.23$). At the firing threshold several graded detectors (331, 534, 1242) have as low a residue recall as the zinc latents; by the residue correlation they are not. We are cautious not to over-interpret these results: the motif analysis rests on five latents and the group comparisons on 17.

\section{Harness Calibration Ladder and Robustness}
\label{sec:appendix_robustness}

At the standard restricted generation budget, the judge scores our 20 ground-truth amino-acid-identity latents at $0.17$ accuracy, indistinguishable from chance. We traced this entirely to a harness artifact (a tight token cap plus a fragile parser), not to the latents. We only built this positive-control panel because a suspiciously flat, near-chance score across every latent in an early pilot made us distrust the harness, not because we anticipated the failure mode in advance. Without it, that at-chance score would have looked identical to genuine uninterpretability.

\begin{figure}[H]
    \centering
    \includegraphics[width=0.72\textwidth]{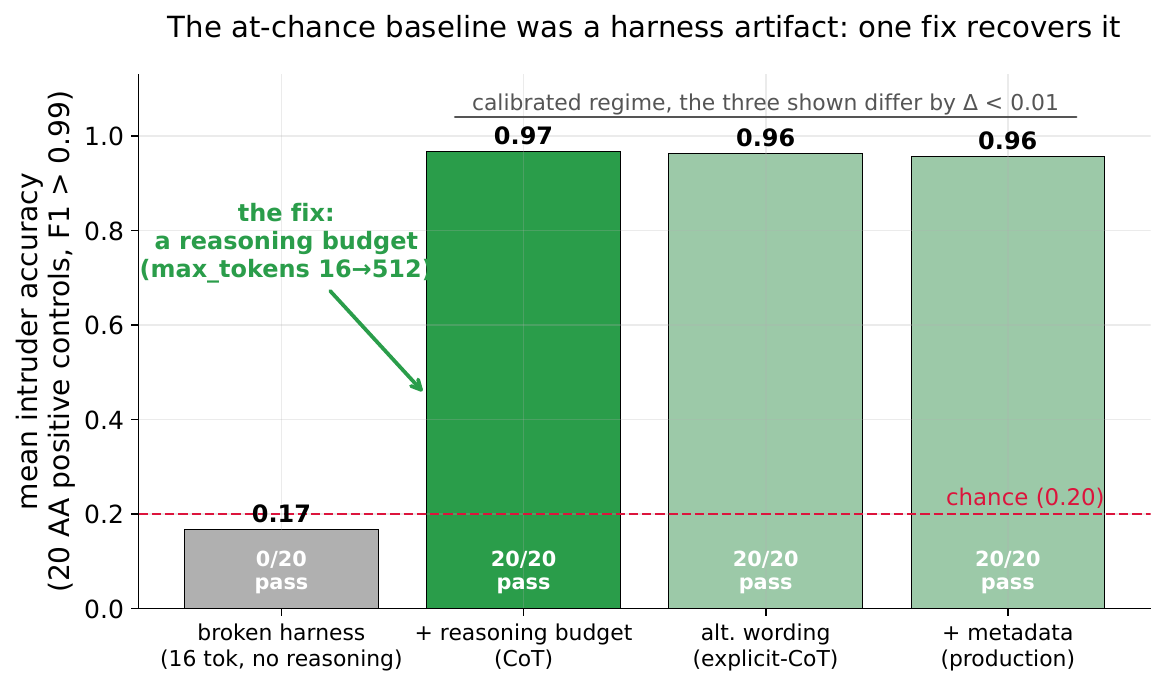}
    \caption{\textbf{Reasoning-budget ablation ladder.} Identical positive-control latents across prompt architectures: an immediate-answer restriction (16 tokens) scores $0.17$, and raising the output-token allowance to 512 tokens recovers $0.97$. Across the three calibrated variants shown, alternative wording ($0.96$) and metadata inclusion ($0.96$) add negligible variance ($\Delta<0.01$). A fourth, independently written paraphrase reaches $0.85$ (Appendix~\ref{sec:appendix_robustness}).}
    \label{fig:fig1_ladder}
\end{figure}

Raising the judge's reasoning and output-token allowance from 16 to 512 tokens recovers accuracy to $0.97$ ($20/20$) and produces a graded rather than binary continuum (Table~\ref{tab:harness_controls}, Fig.~\ref{fig:fig1_ladder}): contextual concepts score $0.48$, random latents $0.29$, and a scrambled negative control sits at chance ($0.17$). The contextual-concept row does not show the judge reading context. These are the best single latents for six structural and SwissProt concepts, and the three that reach an accuracy of $0.5$ (features 249, 297 and 516) also have fairly high residue precision ($0.59$ to $0.67$, on Gly, Ala and Ser), while the other three have $0.15$ to $0.25$ (Section~\ref{sec:judge}).

\begin{table}[H]
\centering
\caption{\textbf{Calibrated intruder detection across control panels (chance $= 0.20$).} Pass rate is the one-sided binomial test against $p_0 = 0.2$ at $\alpha = 0.05$, which is a weaker criterion than the accuracy $\geq 0.5$ bar the discovery sweep uses. The scrambled control is unaffected by the intruder construction of Appendix~\ref{sec:appendix_intruder}, since it draws all five windows from different features and so re-marks nothing.}
\label{tab:harness_controls}
\vskip 0.05in
\small
\begin{tabular}{lccc}
\toprule
\textbf{Feature Group} & \textbf{$n$} & \textbf{Mean Acc.} & \textbf{Pass Rate} \\
\midrule
AA Identity (positive control) & 20 & 0.97 & 20/20 \\
Contextual concepts & 6 & 0.48 & 5/6 \\
Random latents & 13 & 0.29 & 4/13 \\
Scrambled (negative control) & 120 trials & 0.17 & n.s.\ ($p = 0.85$) \\
\bottomrule
\end{tabular}
\end{table}

The recovery is model-agnostic (Claude Sonnet 4.6: $0.96$; Haiku 4.5: $0.97$; Opus 4.8: $0.96$; all $20/20$) and the broken 16-token gate is itself model-dependent; it collapses to chance on Sonnet ($0.17$) and Haiku ($0.13$) but survives on Opus ($0.86$) because Opus tends to emit the bare integer index the fragile parser expects. This reinforces that the fault lived in the harness, not the latents. Across four prompt phrasings (Inline CoT, Explicit CoT, Explicit+Meta, Paraphrased), accuracy holds at $0.85$--$0.97$. The three calibrated production variants sit within $\Delta<0.01$ of each other at $0.957$--$0.967$, and the independently written paraphrase accounts for the lower end at $0.85$, which is the widest any single wording moves the judge. All four remain far above the broken gate at $0.17$, confirming the token allowance, rather than the wording, is the critical variable. These controls are all amino-acid-identity latents, so they show that the harness can see residue identity, not that it can see context (Section~\ref{sec:judge}).

\section{Per-Feature Ground-Truth Validation}
\label{sec:appendix_named_finds}

Table~\ref{tab:named_finds} gives the full precision/recall/fold breakdown behind Section~\ref{sec:verify}, for every named low-frequency find discussed in the main text.

\begin{figure}[h]
    \centering
    \includegraphics[width=0.95\textwidth]{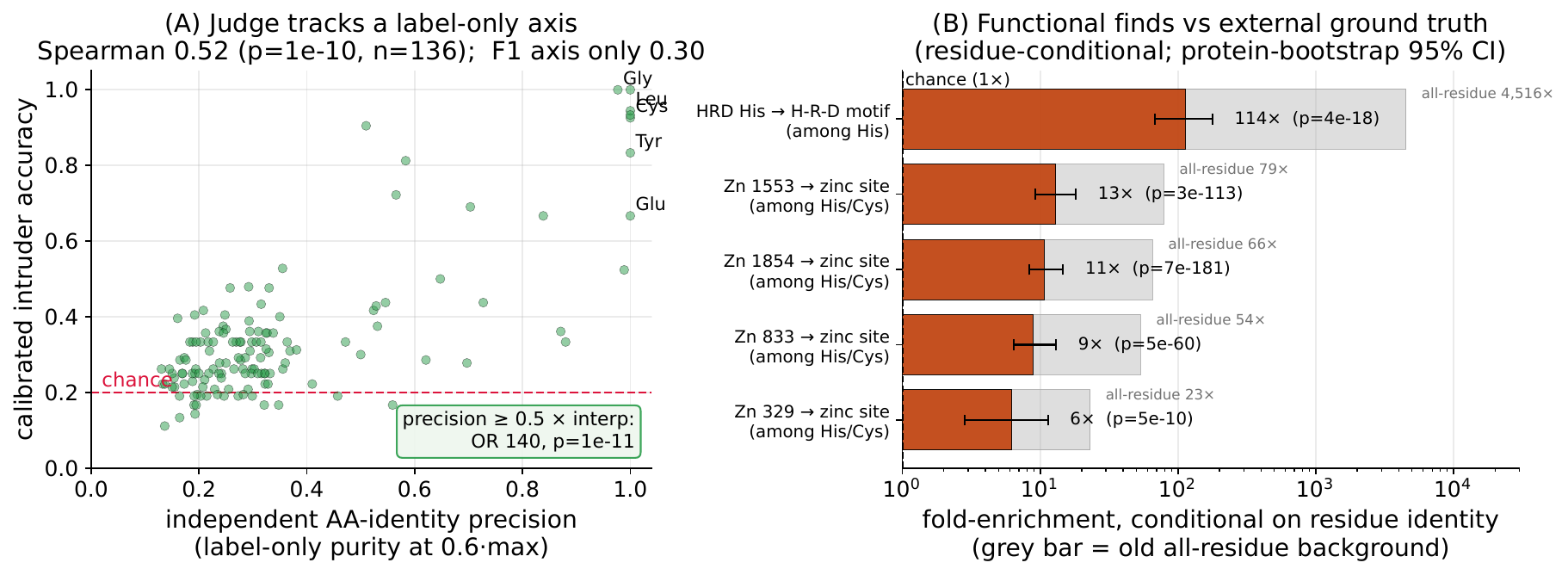}
    \caption{\textbf{What the judge tracks, and ground truth for the motifs.} \textbf{(A)} Calibrated intruder accuracy tracks residue precision, which needs only the sequences and no curated annotations ($\rho = 0.52$); by Section~\ref{sec:judge} this is largely because the judge detects residue identity, so it is not independent validation of the judge; a $2\times2$ split at the residue-precision threshold gives OR $= 140$, with only one of 111 low-precision latents judged interpretable. This analysis covers the 136 of 141 sweep latents that fire on at least 20 residues; the five too-sparse latents excluded here include two of the 17 interpretable finds. \textbf{(B)} Firing enrichment for the metal-coordination and catalytic motifs, against curated SwissProt zinc annotations and, for the catalytic latent, the HRD sequence motif, conditional on residue identity, with a by-protein bootstrap 95\% CI. Grey bars show the all-residue background, which ignores residue identity. The HRD-motif histidine hits the exact His-Arg-Asp motif on nine of the 23 HRD motifs in the corpus and is enriched $114\times$ among histidines, and within HRD-containing proteins it fires only on that histidine. The zinc-coordination latents, including 329 just below the interpretability bar, reach $6$ to $13\times$ among histidines and cysteines.}
    \label{fig:fig5}
\end{figure}

\begin{table}[t]
\centering

\caption{\textbf{Every interpretable latent's auto-interp description, scored against external ground truth.} One row per latent clearing the interpretability bar, ordered by intruder accuracy; $^{*}$ marks feature 329, which falls just short of the bar but is reported alongside the other zinc latents. Each description names a residue class and, in all but one case, a structural or functional \emph{context}; the two halves are scored separately under mappings fixed before any enrichment was computed. Precision, recall and \emph{residue fold} refer to the residue claim, scored against amino-acid identity; they are for the residue class the description names, which can differ from $X$ in Table~\ref{tab:two_kinds} (1553: precision 1.000 on His/Cys, against residue precision 0.58 on His). \emph{Context} precision and fold score the named environment against SwissProt residue annotations or Boltz-DSSP secondary structure, \emph{among residues of the class the description names}, with a by-protein bootstrap 95\% CI; \emph{fails} means that interval does not lie entirely above 1. All four zinc rows use His or Cys as the zinc-coordinating class, although 1854 and 833 name only cysteine; against cysteine alone they give $6.9\times$ and $6.2\times$, with the same verdicts. An all-residue background over-credits a detector that already fires on the right residue: it would put the HRD motif at $4{,}516\times$, the zinc rows at $23$ to $79\times$, isolated-Cys at $68.5\times$ and glycine-in-coil (331) at $1.85\times$. For 1446, 484 and 329, which fire off their named residue 30\%, 35\% and 39\% of the time, context precision and fold use only the on-residue firings. $\dagger$ feature 354 fires on $86.5\%$ of all prolines in the corpus, so no context restriction of it can beat the proline background by more than $1.16\times$.}
\label{tab:named_finds}
\footnotesize
\makebox[\textwidth][c]{
\begin{tabular}{rlccrcll}
\toprule
 & & \multicolumn{3}{c}{Residue claim} & \multicolumn{2}{c}{Context claim} & \\
\cmidrule(lr){3-5} \cmidrule(lr){6-7}
Feat & Auto-interp description & Prec. & Recall & Fold & Prec. & Fold (95\% CI) & Verdict \\
\midrule
331 & Gly in linker/loop regions & 0.977 & 0.097 & $14.6\times$ & 0.68 & $1.13$ ($1.07$--$1.19$) & holds \\
354 & Pro at structural boundaries & 1.000 & 0.865 & $18.2\times$ & 0.29 & $0.94$ ($0.92$--$0.95$) & fails$^\dagger$ \\
1242 & Leu/Ile at C-terminal helix boundary & 1.000 & 0.018 & $7.0\times$ & 0.06 & $0.77$ ($0.51$--$1.06$) & fails \\
1652 & Trp, single or tandem & 1.000 & 0.977 & $84.9\times$ & --- & --- & no category \\
796 & isolated Cys & 1.000 & 0.256 & $53.7\times$ & 0.86 & $1.28$ ($1.21$--$1.35$) & holds \\
1095 & aromatics (Y/F/W/H) near signal peptide & 1.000 & 0.049 & $9.7\times$ & 0.01 & $1.91$ ($0.90$--$3.01$) & fails \\
534 & Tyr at structured/disordered boundary & 1.000 & 0.044 & $35.4\times$ & 0.02 & $1.29$ ($0.27$--$2.76$) & fails \\
1553 & zinc coordination (His/Cys ligands) & 1.000 & 0.020 & $22.8\times$ & 0.70 & $13.0$ ($9.2$--$18.1$) & holds \\
1805 & Ser/Thr immediately post-helix/strand & 0.995 & 0.015 & $7.5\times$ & 0.22 & $1.62$ ($1.32$--$1.92$) & holds \\
1817 & HRD-motif catalytic His & 1.000 & 0.003 & $39.7\times$ & 0.53 & $114$ ($68$--$178$) & holds \\
1499 & Lys/Arg at binding or active site & 1.000 & 0.005 & $8.7\times$ & 0.17 & $20.3$ ($12.8$--$29.6$) & holds \\
1237 & Glu at $\alpha$-helix N-cap & 1.000 & 0.004 & $14.6\times$ & 0.13 & $1.95$ ($0.70$--$3.47$) & fails \\
1854 & zinc coordination (Cys-rich motifs) & 0.839 & 0.081 & $45.0\times$ & 0.58 & $10.8$ ($8.3$--$14.7$) & holds \\
1446 & Lys/Arg clusters in disordered regions & 0.704 & 0.005 & $6.1\times$ & 0.53 & $4.44$ ($3.44$--$5.45$) & holds \\
1448 & Asn at coiled-coil helix boundaries & 1.000 & 0.001 & $23.9\times$ & 0.83 & $33.6$ ($22.0$--$55.9$) & holds \\
833 & Cys in zinc-coordinating motifs & 0.989 & 0.048 & $53.1\times$ & 0.48 & $8.88$ ($6.44$--$13.0$) & holds \\
484 & Gln runs in disordered/low-complexity & 0.648 & 0.008 & $14.9\times$ & 0.31 & $2.36$ ($1.22$--$3.67$) & holds \\
329$^{*}$ & zinc coordination (weakest of the four) & 0.607 & 0.006 & $13.9\times$ & 0.33 & $6.20$ ($2.83$--$11.4$) & holds \\
\bottomrule
\end{tabular}
}
\end{table}

Table~\ref{tab:named_finds} scores both halves of every description: the residue class it names, and the structural or functional context it additionally names. Two of the five context failures (features 534 and 1237) rest on four and seven firing residues inside the claimed context and are underpowered rather than clearly refuted. Five clauses across four descriptions name no available ground-truth category and are recorded as unscored rather than dropped, transmembrane environment being the most common: the annotation set carries no transmembrane track. Note also that the secondary-structure categories derive from DSSP on Boltz-predicted structures and so are not fully model-external; the SwissProt categories and the residue identities are.

\section{Subspace Stability: Full Method and Positive-Control Ladder}
\label{sec:appendix_subspace_stability}

All subspace analyses use the seed-1/2/3 TopK SAEs for Boltz-1 layer 20, recycle 1, over the identical 196,175-residue SwissProt subset. Because SAE inputs are identical across seeds, a target latent's per-residue activation vector is a fixed, seed-independent signature of the region it occupies. Auto-interp runs on seed 1 only; the corresponding feature in seeds 2/3 is located purely by firing-signature match, never re-interpreted. For each seed, the concept's atom group is the set of up to eight latents whose per-residue firing has the largest absolute Pearson correlation ($r \geq 0.15$) with the seed-1 signature. Decoder directions are $L_2$-row-normalized; each group's decoder span is obtained via SVD (singular values $>10^{-8}$); subspace stability is the mean $\cos^2\theta$ over principal angles between $\text{span}(G_1)$ and $\text{span}(G_s)$, $s\in\{2,3\}$, on $[0,1]$, where the random-subspace floor is $\approx d/384$.

\begin{figure}[h]
    \centering
    \includegraphics[width=0.95\textwidth]{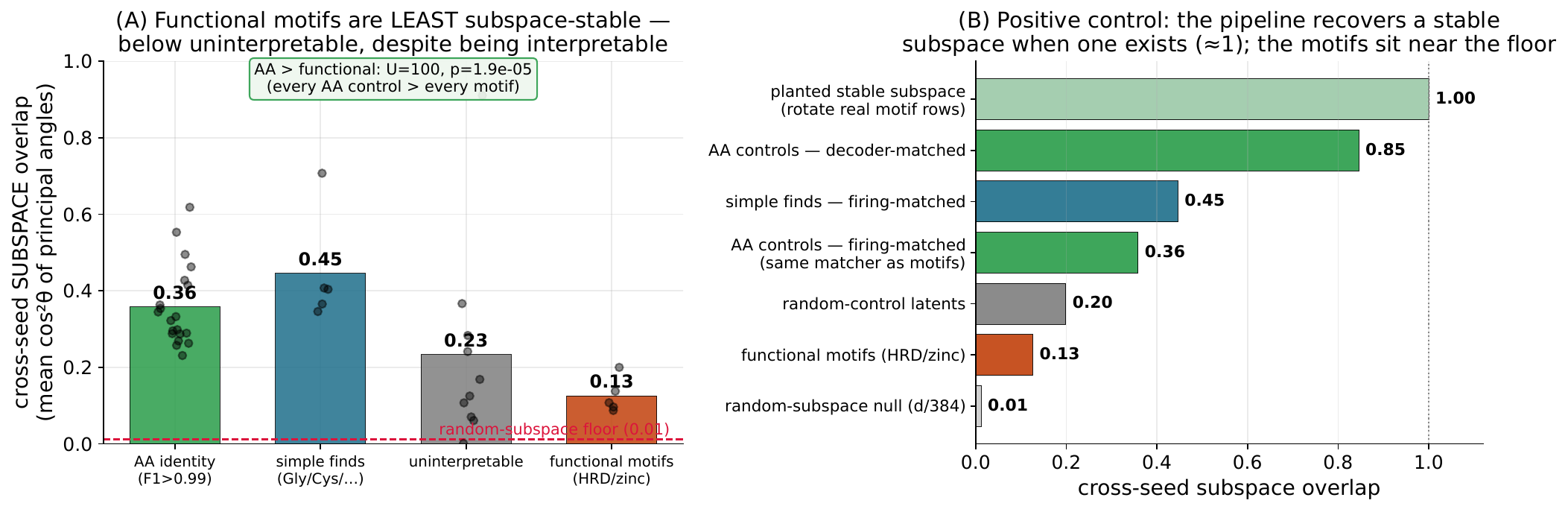}
    \caption{\textbf{Subspace-level stability, and its positive-control ladder.} For each feature, the co-firing decoder group is located independently in all three seeds and compared by mean $\cos^2$ of principal angles (1 = identical subspace). \textbf{(A)} The five metal-coordination and catalytic motifs (0.13) score no higher than uninterpretable latents (mean 0.23, raised by one latent at 0.91; median 0.15) and far below amino-acid controls (0.36) run through the identical pipeline. \textbf{(B)} Positive-control ladder: a planted rotated subspace (1.00) and a genuinely stable atom group (0.85) confirm the metric has range and the motifs sit near the floor, not near either.}
    \label{fig:figS2_manifold_subspaces}
\end{figure}

\textbf{Positive-control ladder.} (1) \emph{Metric unit test}: a synthetic rotated basis of a fixed plane returns $1.00$; a random $d$-flat returns $\approx d/384$. (2) \emph{Rotation invariance}: on real functional decoder rows, identity overlap is $1.00$ and rotating within the same span returns $1.00$; a planted same-subspace partner with Gaussian noise degrades smoothly. (3) \emph{Stable-atom recovery}: the 20 amino-acid controls recover high subspace overlap under both decoder-cosine (Hungarian, $0.85$) and firing-correlation (the same matcher used for the unsupervised finds, $0.36$) matching, and the simple structural discoveries (Gly, Cys, Tyr, Leu/Ile, Glu) are themselves reproducible at $0.45$. The fair comparator for these motifs is therefore the firing-matched amino-acid controls ($0.36$), since the motifs have no stable decoder axis to match by construction and even against that comparator they fall to $0.13$ (Fig.~\ref{fig:figS2_manifold_subspaces}). Overlap does not track activation frequency across the panel, ruling out a sparsity-of-estimation explanation. These controls show that the metric recognises a shared subspace when one is supplied; we did not vary the group size or the correlation threshold, so the motifs' low score rules out a shared span under this grouping rather than under every grouping.

\section{Prompts}
\label{sec:appendix_prompts}

All prompts are reproduced verbatim from \texttt{autointerp/intruder.py} and \texttt{autointerp/describe.py}, with the example count resolved to the value used (four genuine examples plus one intruder). Long lines are wrapped here for width; the prompts sent to the model are unwrapped.

\textbf{Intruder judge, system prompt} (production configuration, with metadata; the annotation glossary at the end of this appendix is appended):
{\footnotesize
\begin{verbatim}
You are analyzing what a protein-model feature detects. In each example, the exact
residues the feature fired on are listed in highlighted_residues=[...] and also wrapped
in <<< >>> in the window.

You will see 5 numbered protein examples. In 4 of them the highlighted residues share a
common pattern -- this may be as simple as the same amino-acid IDENTITY, or a shared
biochemical type (charge, hydrophobicity, size), or a shared local sequence/structural
context. One of them -- the INTRUDER -- does not share it.

You are NOT told the pattern. Reason briefly: note what the highlighted residues have in
common across the majority (start from the SIMPLEST explanation -- the literal amino-acid
identity -- before anything more complex), then identify the one that breaks it.

End your response with a final line exactly: ANSWER: <index 1-5>.
\end{verbatim}
}

\textbf{Intruder judge, user message.} Each example is rendered on one line. A real example from feature 1817 (protein name shortened, line wrapped at the field separators):
{\footnotesize
\begin{verbatim}
highlighted_residues=[H]  |  Serine/threonine-protein kinase PAK mbt (...) |
window: KQCLKALAYLHSQGVI<<<H>>>RDIKSDSILLAADGRV |
nearby: Active site (487-487): active_site:proton_acceptor | note=Proton acceptor;
Domain (368-619): domain:protein_kinase | note=Protein kinase
\end{verbatim}
}
\noindent The code adds a \texttt{structure(H/E/C)} field when a secondary-structure track is attached, which was never the case in this study. The five examples follow a line reading \texttt{Examples:}, are numbered 1 to 5, and are followed by:
{\footnotesize
\begin{verbatim}
Which numbered example is the intruder? Reason briefly, then end with a final line
exactly: ANSWER: <index 1-5>.
\end{verbatim}
}

\textbf{Describe pass, system prompt} (the annotation glossary below is appended):
{\footnotesize
\begin{verbatim}
You are an expert computational biologist and protein biochemist.

Your task is to infer and clearly describe what a given protein-model SAE feature is
detecting, based on protein examples where it activates. Each example shows a residue
window around the feature's activation peak; residues where the feature fires are
wrapped in <<< >>>. Examples are grouped by activation strength (HIGH/MEDIUM/LOW), and
nearby SwissProt annotations are provided.

ANALYSIS PROCESS (work through these mentally before answering):
1. PATTERN IDENTIFICATION - What amino acids, motifs, or properties (charge,
   hydrophobicity, size, aromaticity) appear at the highlighted positions?
2. COMPARATIVE ANALYSIS - What distinguishes HIGH from MEDIUM from LOW activation? Which
   annotations co-occur with strong activation?
3. HYPOTHESIS GENERATION - 2-3 distinct hypotheses (sequence motif, structural context,
   functional role, protein family).
4. HYPOTHESIS EVALUATION - Which hypothesis best explains ALL the data, including
   outliers?
5. BOUNDARIES - What would this pattern fail to predict?

QUALITY STANDARDS:
- Use only the information provided. Prefer mechanistic explanations over surface
  observations.
- Make the description useful for predicting activation on new, unseen proteins.
- Note surprising or tentative findings explicitly.

Output format: respond with a single JSON object with exactly two string fields:
- "short_description": exactly 1 sentence, MUST start with "The feature activates on".
- "long_description": 3-8 sentences on motifs, regions, structural/functional patterns,
  plausible mechanisms, and caveats. MUST start with "The activation patterns are
  characterized by:".
Do not include any other fields.
\end{verbatim}
}

\textbf{Describe pass, user message.} After a header line, \texttt{Feature <index>. Activating examples (grouped by strength):}, windows are listed strongest first, each as
{\footnotesize
\begin{verbatim}
=== <protein id> | <HIGH/MEDIUM/LOW> (activation <0.00-1.00>) ===
name: <protein name>
window @ residue <peak position> (firing residues in <<< >>>): <window>
nearby SwissProt annotations: <up to six annotations>
\end{verbatim}
}
\noindent followed by \texttt{Return ONLY a JSON object with "short\_description" and "long\_description".} As for the judge, a secondary-structure line is added only when a track is attached, which it never was here.

\textbf{Annotation glossary} (appended to both system prompts):
{\footnotesize
\begin{verbatim}
SwissProt annotation glossary (positions are 1-indexed residue ranges):
- active_site / binding_site: catalytic residues / ligand- or metal-binding residues.
- domain / region / motif / repeat: a structural or functional segment;
  region:disordered = intrinsically disordered (no fixed 3D structure).
- compositional_bias: a stretch enriched in particular residues (acidic = D/E;
  basic = K/R/H; "low complexity").
- modified_residue / lipidation / glycosylation / disulfide_bond: post-translational
  modifications or crosslinks.
- helix / beta_strand / turn / coiled_coil, and a "secondary structure (H/E/C)" track:
  local 3D conformation (H = helix, E = beta-strand, C = coil/loop).
- signal_peptide / transit_peptide: cleaved N-terminal targeting sequences.
- zinc_finger: a zinc-coordinating module.
\end{verbatim}
}

\newpage

\end{document}